\documentclass[prd,twocolumn,,floatfix,amsmath,amssymb]{revtex4-1}
\usepackage{amsmath,amsfonts,amsthm,amssymb,slashed,graphicx}
\usepackage{xcolor}
\usepackage{graphicx}
\usepackage{epstopdf}
\usepackage{array,booktabs}
\usepackage{hyperref}
\usepackage{bbm} % for \mathbbm{1}
\usepackage{mathtools} % for \mathclap
\usepackage{tabulary}
\usepackage{multirow}
\usepackage{array,arydshln}
\hypersetup{linktoc = all}  %hyperref settings
\hypersetup{pdfborderstyle={/S/U/W 0.5}}
\usepackage{tikz}
\usepackage{tkz-tab}
\usepackage{caption}
\usepackage{subcaption}
\usetikzlibrary{shapes.geometric}
\usetikzlibrary{intersections}
\usetikzlibrary{calc}
\usetikzlibrary{decorations.pathmorphing}
\newcounter{mycnt}

\begin{document}

\title{Geons and the Quantum Information Metric}

\author{Musema Sinamuli}
\email{cmusema@perimeterinstitute.ca}
\affiliation{Department of Physics and Astronomy, University of Waterloo, Waterloo, Ontario N2L 3G1, Canada}
\affiliation{Perimeter Institute for Theoretical Physics, Waterloo, Ontario, N2L 2Y5, Canada}

\author{Robert B. Mann}
\email{rbmann@uwaterloo.ca}
\affiliation{Department of Physics and Astronomy, University of Waterloo, Waterloo, Ontario N2L 3G1, Canada}

\begin{abstract}
We investigate the proposed duality between a quantum information metric in a $\mbox{CFT}_{d+1}$ and the volume of a maximum time slice in the dual $\mbox{AdS}_{d+2}$ for topological  geons. 
Examining the specific cases of BTZ black holes and planar Schwarzschild-AdS black holes, along
with their geon counterparts, we find that the proposed duality relation for  geons  is the same apart
from a factor of 4.  The information metric therefore provides a probe of the topology of the bulk spacetime.  
\end{abstract}

\maketitle
\section{Introduction}

Research  in holography has been of interest since the advent of the 
 AdS/CFT correspondence conjecture  \cite{thelarge}, and has since been extended to more general 
 notions of gauge/gravity duality.  A recent new example is the proposal 
 that there exists a dual connection between the geometric length of an Einstein-Rosen bridge (ERB) and the computational complexity  of the dual CFT's quantum states \cite{computationalcom, addendum, entanglement}. This in turn has led to a broad number of investigations on the topic, many concerned with its quantum informational aspects and whether or not there exists a 
CFT quantity that  is dual to a volume of a co-dimension one time slice in AdS, analogous to the relationship between  holographic entanglement entropy and the area of codimension two extremal surfaces \cite{Ryu}.  
 
 A recent proposal to this end has been that of a correspondence between  a quantum information quantity referred to as the {\it information metric} (or the {\it fidelity susceptibility}) and the volume of a maximum time slice of an AdS-like black hole \cite{gravitydual}. The notion of a quantum information metric $G_{\lambda\lambda}$ has been around for quite some time and consists of the comparison between a quantum state of the CFT and its counterpart in a {\it marginal} deformation thereof, giving rise to a term proportional to the fidelity susceptibility. The proposed corresponding bulk quantity is  the maximal volume of a time slice connecting the two boundaries (CFTs) of the dual AdS black hole.  After comparing the computations of each, the following relation 
\begin{equation}
\label{intro1}
G_{\lambda\lambda}=n_d\frac{\mbox{Vol}(\Sigma_{max})}{R^{d+1}}
\end{equation}
was proposed \cite{gravitydual},
where $n_d$ is ${\cal{O}}(1)$ constant, $\Sigma$ a time slice, $d$ the dimension of the spacetime and $R$ the radius of the AdS spacetime.

Our intention here is to explore the proposal \eqref{intro1} for topologically non-trivial spacetimes. We specifically shall consider geons.  Introduced by Wheeler in 1955 as bound gravitational and electromagnetic entities, geons were found to be unstable due to the tendency of a massless field to either disperse to infinity or collapse into a black hole.  Topological geons were introduced somewhat later \cite{introduction}, generalizing the original construction by allowing nontrivial spatial topology with a black hole horizon.   These objects  provide an arena for advancing our understanding of black holes in both classical and quantum contexts and so are of considerable interest.  Specific examples include topological censorship theorems
\cite{FriedSchWi}, Hawking radiation \cite{louko-marolf-rp3,langlois-rp3}, and the behaviour of Unruh de Witt detectors as probes of hidden topology \cite{lookinginside}. The topological identification inherent in the construction of geons affects both the bulk interior and its posited relationship to the dual CFT, making these interesting objects of study insofar as understanding the proposal  \eqref{intro1} is concerned. 
 
We consider the corresponding geon space of two distinct black holes:  the BTZ black hole in $d=3$ and the AdS-Schwarzchild black  brane in $(d+2)$ dimensions. 
 We compute the information metric and the maximal volume of a time slice in the bulk for each case and compare them. Our results suggest that the coefficient $n_d$ is a function of spacetime topology:  we find in both cases that the proportionality in \eqref{intro1} is preserved, but the coefficient is increased by a factor of 4. The information metric is indeed sensitive to spacetime topology.

Our paper is organized as follows. In section 2 we briefly review the notion and construction of geons. In section 3, we describe the information metric and compute it for the black holes as well as for their geon counterparts. Section 4 consists of a computation of the maximal volume of a spatial slice for each case, and then extends these considerations to the planar Schwarzschild metric and its geon counterpart.
In the concluding section we summarize our results and their  implications. An appendix contains some computational details of our work.

\section{Review of geons}
 
The construction of a geon generally makes use of a freely acting involutive isometry that acts
on a (black hole) spacetime.
 A necessary condition for their construction is that  the spacetime must be time orientable and foliated by spacelike hypersurfaces with single asymptotic region  \cite{geonswith}.  The asymptotic region is generally also required to be stationary and allow conserved charges to be defined by appropriate integrals. 

Let us illustrate this for the BTZ black hole, which is a quotient of   $\mbox{AdS}_3$, and its geon counterpart (the $\mbox{RP}_2$ geon) is obtained via further quotienting  as follows. The metric for the non-rotating BTZ black hole is \cite{lookinginside}
\begin{eqnarray}
\label{ads3}
ds^2&=&-f(r)dt^2+dr^2/f(r)+r^2d\phi^2\nonumber\\
f(r)&=&-M+r^2/l^2 ~~~~~~~~~~r_+=l\sqrt{M}.
\end{eqnarray}
with ~$l$ the AdS radius and $r_+$ the horizon radius. Writing equation (\ref {ads3}) in Kruskal coordinates, it takes the form 
\begin{equation}
ds^2=-\frac{l^2}{1+UV}\big[-4dUdV+M(1-UV)^2d\phi^2\big]
\end{equation}
and we remark that the metric in the new coordinates is invariant under the interchange of $U$ and $V$ as well as under translations of $\phi$.  The geon   is the resulting space obtained via the identification
\begin{equation}
\label{geonspace}
J: (U,V,\phi)\rightarrow (V,U,P(\phi))
\end{equation}
where
\begin{equation}
  \Gamma=\left\{Id, J\right\}\simeq \mathbb{Z}_2
 \end{equation}
 is the group generated by the freely acting involutive isometry $J$.  Equation \eqref{geonspace} 
 corresponds  to 
\begin{equation}
\label{geontransf}
(t,\phi)\rightarrow (-t,\phi+\pi)
\end{equation}  
in the original coordinates \eqref{ads3}.

\section {2+1 Geon information metric}

 The proposed  quantum information metric $G_{\lambda\lambda}$  \cite{gravitydual} is defined by considering  the fidelity susceptibility  between neighbouring states
\begin{equation}\label{infomet}
|\langle\psi(\lambda)|\psi(\lambda+\delta\lambda)\rangle|=1-G_{\lambda\lambda}(\delta\lambda)^2+O((\delta\lambda)^3)
\end{equation}
in the dual CFT, where the parameter $\lambda$ generates a 
  one parameter family of states $|\psi (\lambda)\rangle$.  
In this section we shall compute this quantity for the  states associated with the geon dual, considering only the time-dependent states. These states in the thermo-field double (TFD) description of finite temperature state in a $2d$ CFT dual to the $\mbox{AdS}_3$, read  
\begin{equation}
\label{tfd}
|\Psi_{TFD}\rangle\propto e^{-i(H_1+H_2)t}\sum_{n}e^{-\frac{\beta}{4}(H_1+H_2)}|n\rangle_1|n\rangle_2
\end{equation}
where $H_{1,2}$ are the Hamiltonians of the $\mbox{CFT}_{1,2}$ respectively, and $|n\rangle_{1,2}$ are the unit norm eigenstates of the $\mbox{CFT}_{1,2}$ respectively. 
 It is important to underline that these Hamiltonians are identical.

We shall work with the Euclidean path-integral formalism of the $2d$ CFT, in which case  we must compute $\langle\Psi^{'}_{TFD}(\tau)|\Psi_{TFD}(\tau)\rangle$, where
\begin{eqnarray}
|\Psi_{TFD}(\tau)\rangle&=&\sum_{n}e^{-(\frac{\beta}{4}+\tau)(H_1+H_2)}|n\rangle_1|n\rangle_2
\nonumber \\
|\Psi^{'}_{TFD}(\tau)\rangle&=&\frac{1}{N}\sum_{n}e^{-(\frac{\beta}{4}+\tau+\epsilon)(H^{'}_1+H^{'}_2)}|n\rangle_1|n\rangle_2
\end{eqnarray}
in terms of the Euclidean time.  Here
 $N$ is a normalization factor and the state $|\Psi^{'}_{TFD}(t)\rangle$ being an eigenstate of 
the Hamiltonian $H^{'}_1+H^{'}_2$, which is an  infinitesimal marginal deformation of the original  Hamiltonians.
Following the construction in \cite{gravitydual}, the scalar product \eqref{infomet} takes the form 
\begin{equation}\label{scalprod}
\langle\Psi^{'}|\Psi\rangle=\frac{\langle \exp [-\int^{\frac{3\beta}{4}-\tau-\epsilon}_{\frac{\beta}{4}+\tau+\epsilon}d\tau_1\int d^dx\delta L]\rangle}{\big[\langle \exp [-(\int^{\frac{3\beta}{4}-\tau-\epsilon}_{\frac{\beta}{4}+\tau+\epsilon}+
\int^{\frac{\beta}{4}+\tau-\epsilon}_{-\frac{\beta}{4}-\tau+\epsilon})d\tau_1\int d^dx\delta L]\rangle\big]^{\frac{1}{2}}}
\end{equation}
with $\delta L= L^{'}-L\equiv \delta\lambda {\cal{O}}(\tau,x)$.  $L$ and $L^{'}$ are Lagrangian densities associated the the Hamiltonian $H$ and $H^{'}$ respectively, and we shall henceforth assume the perturbation $\delta\lambda\cdot {\cal{O}}(\tau , x)$ is marginal (of dimension
$\Delta=d+1=2$ for the present case).
$\epsilon$ is a very small parameter that we regard as a cut-off.
 
 In the Euclidean path-integral formalism
 the two point function for the BTZ black hole is defined on $S^1\times S^1$ (where one $S^1$ is the thermal circle with period $\beta$) and takes the form \cite{eternalblack, ongeodesic}
\begin{eqnarray}
\label{2pointBTZ}
&&\langle{\cal{O}}(\phi_1,\tau_1){\cal{O}}(\phi_2,\tau_2)\rangle_{BTZ}\nonumber\\
&=&\sum_{n}\frac{(\frac{\pi}{\beta})^{2\Delta}}{\bigg[\sinh^2\bigg(\frac{\pi(\phi_2-\phi_1+2\pi n)}{\beta}\bigg)+\sin^2\bigg(\frac{\pi(\tau_2-\tau_1)}{\beta}\bigg)\bigg]^\Delta}\nonumber\\
&&
\end{eqnarray}
where~~$\Delta$ is the total conformal dimension of the primary field ${\cal{O}}$.  Upon expanding
\eqref{scalprod} and comparing to \eqref{infomet} we find
\begin{eqnarray}
\label{infometric}
&G^{BTZ}_{\lambda\lambda}&
=\frac{1}{2}\int_{0}^{2\pi}d\phi_1 d\phi_2\int_{\frac{\beta}{4}+\tau+\epsilon}^{\frac{3\beta}{4}-\tau-\epsilon}d\tau_2\int_{-\frac{\beta}{4}-\tau+\epsilon}^{\frac{\beta}{4}+\tau-\epsilon}d\tau_1\nonumber\\
&&\quad \times\langle{\cal{O}}(\phi_1,\tau_1){\cal{O}}(\phi_2,\tau_2)\rangle_{BTZ}
\end{eqnarray}
Noting the identity
\begin{eqnarray}
\label{identity}
\sum_{n}\int_{0}^{2\pi}d\phi_2 f(\phi_2+2\pi n)&=&\sum_{n}\int_{2\pi n}^{2\pi(n+1)}dx_nf(x_n)\nonumber\\
&=&\int_{-\infty}^{\infty}dx f(x)
\end{eqnarray}
with~~~$x_n=\phi_2+2\pi n$,~~~$n$ integer,  we can rewrite \eqref{infometric} as
\begin{eqnarray}
\label{infometric1}
&&G^{BTZ}_{\lambda\lambda}
=\frac{1}{2}\int^{2\pi}_{0}dx_1\int_{-\infty}^{\infty} dx_2\nonumber\\
&& \times \int_{\frac{\beta}{4}+\tau+\epsilon}^{\frac{3\beta}{4}-\tau-\epsilon}d\tau_2\int_{-\frac{\beta}{4}-\tau+\epsilon}^{\frac{\beta}{4}+\tau-\epsilon}d\tau_1
\langle{\cal{O}}(x_1,\tau_1){\cal{O}}(x_2,\tau_2)\rangle
\end{eqnarray}
where $x_1\in [0, 2\pi)$,  $x_2\in\mathbb{R}$ and the new two point function reads  
\begin{eqnarray}
\label{2point}
&&\langle{\cal{O}}(x_1,\tau_1){\cal{O}}(x_2,\tau_2)\rangle \nonumber\\
&& = \frac{(\frac{\pi}{\beta})^{2\Delta}}{\bigg[\sinh^2\bigg(\frac{\pi(x_2-x_1)}{\beta}\bigg)+\sin^2\bigg(\frac{\pi(\tau_2-\tau_1)}{\beta}\bigg)\bigg]^\Delta}
\end{eqnarray}

Now that we have assembled all the ingredients to compute the information metric \eqref{2point} , we define $u=\frac{\pi(\tau_1-\tau_2)}{\beta}$ and likewise $\tilde{x} = \frac{\pi(x_2-x_1)}{\beta}$, noting
the restriction  $(0\leq u\leq\pi)$.  We obtain for the integration over $x_2$ of the integrand in \eqref{infometric1} an integral of the form \cite{gravitydual}
\begin{eqnarray}
\label{integraltrick}
&&\int_{-\infty}^{\infty}d\tilde{x} ~\big(\sinh^2 \tilde{x}+\sin^2 u\big)^{-2}\nonumber\\
&=&\frac{1}{\sin^2 u\cos^2 u}+(u-\pi /2)\frac{2\sin^2 u-1}{\sin^3 u\cos^3 u}
\end{eqnarray}
To find the information metric we have to integrate over all the variables in \eqref{infometric1}. Upon integration of (\ref {integraltrick}) with respect to $\tau_1$, we find 
\begin{eqnarray} 
&&G^{BTZ}_{\lambda\lambda} = - \frac{1}{2}\int^{2\pi}_{0}dx_1 \nonumber \\
&&\quad \times\int_{\frac{\beta}{4}+\tau+\epsilon}^{\frac{3\beta}{4}-\tau-\epsilon}d\tau_2 \left(
\cot 2u+ 2\frac{(u-\pi/2)}{\sin^2 2u} \right)\left|^{{}^{\frac{\pi}{\beta}(\beta/4+\tau-\epsilon-\tau_2)}}_{{}_{-\frac{\pi}{\beta}(\beta/4+\tau-\epsilon+\tau_2)}} \right.\nonumber \\
&&\qquad =\frac{\pi V_1}{8\epsilon}-\frac{\pi V_1}{2\beta}+\frac{2\pi^2 V_1}{\beta^2}\tau\cot\big(\frac{4\pi\tau}{\beta}\big)
\label{G-btz}
\end{eqnarray}
where $V_1= 2\pi$ is the finite volume obtained upon integration over $x_1$ from $0$ to $2\pi$. 

For the geon the corresponding calculation is similar. The 2 point function is 
\cite{ongeodesic, behind}
\begin{equation}
\label{geoncont}
\langle{\cal{O}}(x){\cal{O}}(x^{'})\rangle_{geon}=\langle{\cal{O}}(x){\cal{O}}(x^{'})\rangle_{BTZ}+\langle{\cal{O}}(x){\cal{O}}(Jx^{'})\rangle_{BTZ}
\end{equation}
where the first term  is the contribution \eqref{2pointBTZ} that appears in the BTZ case and the second term is the geon contribution to the two point function. Using  (\ref {geontransf}) it reads 
 \begin{eqnarray}
 \label{geon2point}
&&\langle{\cal{O}}(\phi_1,\tau_1){\cal{O}}(J\phi_2,J\tau_2)\rangle_{BTZ}\nonumber\\
&=&\sum_{n}\frac{(\frac{\pi}{\beta})^{2\Delta}}{\bigg[\sinh^2\bigg(\frac{\pi(\phi_2-\phi_1+\pi+2\pi n)}{\beta}\bigg)+\sin^2\bigg(\frac{\pi(\tau_2+\tau_1)}{\beta}\bigg)\bigg]^\Delta}\nonumber\\
&&
\end{eqnarray}
The information metric for the geon will be ${G}^{geon}_{\lambda\lambda} = {G}^{BTZ}_{\lambda\lambda} + \tilde{G}^{BTZ}_{\lambda\lambda}$, the latter contribution coming from \eqref{geon2point}. Making use of the identity \eqref{identity}, we obtain from this term an integral of the form  \eqref{infometric1} but where 
\begin{eqnarray}
\label{2pointgeon}
&&\langle{{\cal{O}}(x_1,\tau_1)\cal{O}}(x_2,\tau_2)\rangle \nonumber\\
&=& \frac{(\frac{\pi}{\beta})^{2\Delta}}{\bigg[\sinh^2\bigg(\frac{\pi(x_2-x_1)}{\beta}\bigg)+\sin^2\bigg(\frac{\pi(\tau_2+\tau_1)}{\beta}\bigg)\bigg]^\Delta}\nonumber\\
&&
\end{eqnarray}
The integration over the second term in \eqref{geon2point} proceeds as before, the only distinction being the sign of $\tau_1$.  We obtain
\begin{equation}
\tilde{G}^{BTZ}_{\lambda\lambda}= \frac{\pi V_1}{8\epsilon}-\frac{\pi V_1}{2\beta}+\frac{2\pi^2 V_1}{\beta^2}\tau\cot\big(\frac{4\pi\tau}{\beta}\big)   
\end{equation}
which is  the same result as in \eqref{G-btz}.

 We see that  the quantum information metric for the geon is the sum of these two contributions and so is double of that of the original black hole
\begin{equation}
G^{geon}_{\lambda\lambda}=\frac{\pi V_1}{4\epsilon}-\frac{\pi V_1}{\beta}+\frac{4\pi^2 V_1}{\beta^2}\tau\cot\big(\frac{4\pi\tau}{\beta}\big)
\end{equation}
Returning to the original  (non-Euclidean) coordinates via $\tau=it$, the quantum information metric becomes
\begin{equation}\label{Geon-infomet}
G^{geon}_{\lambda\lambda}=\frac{\pi V_1}{4\epsilon}-\frac{\pi V_1}{\beta}+\frac{4\pi^2 V_1}{\beta^2}t\coth\big(\frac{4\pi t}{\beta}\big)
\end{equation}
In the late time limit $t\gg\beta$, it reduces to
\begin{equation}
G^{geon}_{\lambda\lambda}\simeq\frac{\pi V_1}{4\epsilon}+\frac{4\pi^2 V_1}{\beta^2}t
\end{equation}
In the early time limit $t\rightarrow 0$, we obtain
\begin{equation}
G^{geon}_{\lambda\lambda}\simeq\frac{\pi V_1}{4\epsilon}+\frac{16\pi^3}{3\beta^3}V_1t^2
\end{equation}

We now consider the bulk side of this calculation.
 As prescribed in \cite{locally, openandclose, holographic, holographicdual, aspectsof} a perturbation of the parameter at the time slice $\tau=0$ in the CFT is equivalent to adding a defect brane action
\begin{equation}
S=T\int_{\Sigma}\sqrt{g}
\end{equation}
to the Einstein-Hilbert action. The above statement can also be generalised to time-dependent states. In the case of an infinitesimally small deformation, the quantity $T$ can be approximated to \cite{gravitydual}
\begin{equation}
T\simeq n_d\frac{(\delta\lambda)^2}{R^{d+1}}
\end{equation}
with~ $d=1$ for a $3d$ bulk and $n_d$ an $O(1)$ constant fixed when normalizing the two point function.

%The next steps will consist in the discussion two cases of black holes namely the BTZ black hole and the AdS Schwarzschild black hole, all in $3d$.   

Consider the BTZ black hole in the coordinates   \cite{timeevolution}  
\begin{equation}\label{btz-rho}
ds^2=R^2\big(-\sinh^2\rho dt^2+d\rho^2+\cosh^2\rho dx^2\big)
\end{equation}
which can be obtained from \eqref{ads3} by setting $r = l\sqrt{M}\cosh\rho$ and appropriately rescaling
$t$ and $x$. Here we identify $x\rightarrow x+2\pi$
and so do not have to unwrap the metric as in \cite{gravitydual, timeevolution}.  
 
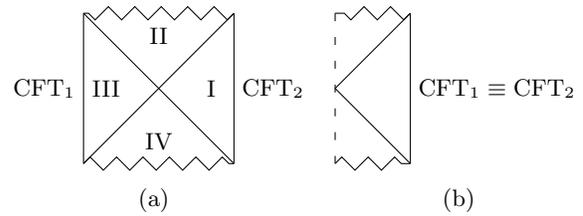
\begin{figure}
\begin{subfigure}[b]{0.22\textwidth}
\centering
\begin{tikzpicture}[scale=1]
\node (I)    at ( 1.7,1)   {I};
\node (I)    at ( 1,1.7)   {II};
\node (I)    at ( 0.3,1)   {III};
\node (I)    at ( 1,0.3)   {IV};
\draw[decorate,decoration=zigzag](0,0)--(2,0);
\draw(2,0)--(2,2)
node[midway, right, inner sep=1mm] {$\mbox{CFT}_2$};
\draw[decorate,decoration=zigzag](2,2)--(0,2);
\draw(0,2)--(0,0)
node[midway, left, inner sep=1mm] {$\mbox{CFT}_1$};
\draw(0,0)--(2,2);
\draw (2,0)--(0,2);
\end{tikzpicture}
\caption{}\label{fig:M1}
\end{subfigure}
\begin{subfigure}[b]{0.22\textwidth}
\centering
\begin{tikzpicture}[scale=1]
\draw[decorate,decoration=zigzag](1,0)--(2,0);
\draw(2,0)--(2,2)
node[midway, right, inner sep=1mm] {$\mbox{CFT}_1\equiv \mbox{CFT}_2$};
\draw[decorate,decoration=zigzag](2,2)--(1,2);
\draw[loosely dashed](1,2)--(1,0);
\draw(1,1)--(2,2);
\draw (1,1)--(2,0);
\end{tikzpicture}
\caption{}\label{fig:M2}
\end{subfigure}
\caption{(a) Penrose diagram of the AdS (BTZ) black hole. The diagram shows  two $2d$ CFTs each at one boundary of the AdS (BTZ) black hole as preconises the AdS/CFT correspondance.(b) Penrose diagram of the AdS (BTZ) geon. This diagram appears to be the half of the AdS (BTZ) black hole one. In fact, the geon space is obtain via the identification (\ref {geonspace}) and  consequently  `splits' the original space into two pieces which are equivalent by a mirror symmetry. As the resulting space happens to be a one-sided black hole, an important and remarkable feature of this diagram is that the two CFTs are now identified and lie on the only remaining boundary}
\label{fig:M3}
\end{figure}

In the region (II) of the Penrose diagram of the black hole \ref{fig:M1},  we use the analytic continuation through the parametrization  $\kappa=-i\rho$~ and~ $\tilde{t}=t+i\pi/2$ and define $\Sigma$ to be the space characterized by $\kappa=\kappa(\tilde{t})$ and such that
\begin{equation}
\mbox{Vol}^{BTZ}(\Sigma)=R^2V_1\int d\tilde{t}\cos\kappa\sqrt{\sin^2\kappa-(\partial\kappa/\partial\tilde{t})^2}
\end{equation}
Denoting by $\kappa_{\ast}$ the value of $\kappa$ for which $\partial\kappa/\partial\tilde{t}=0$ (defined within the interval $(0\leq\kappa_{\ast}<\pi/4)$), and noting that $\dot\kappa
\frac{\partial L}{\partial \dot\kappa} - L$ is a constant of the motion, we can
 repeat the steps  in  \cite{Hartman:2013qma}
 to obtain a maximum volume $\mbox{Vol}(\Sigma_{max})$. This result can be extended to the region (I) in \ref{fig:M1} and we find
 \begin{eqnarray}
\frac{\mbox{Vol}^{BTZ}(\Sigma)}{R^{2}V_1}&=&2\int_{0}^{\kappa_{\ast}}\frac{\cos\kappa d\kappa}{\sqrt{\sin^2(2\kappa_{\ast})/\sin^2(2\kappa)-1}}\nonumber\\
&+&2\int_{0}^{\rho_{\infty}}\frac{\cosh\rho d\rho}{\sqrt{1+\sin^2(2\kappa_{\ast})/\sinh^2(2\rho)}}
\label{volbtz}
\end{eqnarray}
and for the $t$ coordinate  in region (I) 
\begin{eqnarray}
t&=&\int_{0}^{\kappa_{\ast}}\frac{d\kappa}{\sin\kappa\sqrt{1-\sin^2(2\kappa)/\sin^2(2\kappa_{\ast})}}\nonumber\\
&-&\int^{\rho_\infty}_{0}\frac{d\rho}{\sinh\rho\sqrt{1+\sinh^2(2\rho)/\sin^2(2\kappa_{\ast})}}
\end{eqnarray}
 with the integration contour shown in \ref{fig:M4}, and where the factor of 2 comes from the symmetry
in figure \ref{fig:M1}. 
\vskip 5pt  If we define the UV cut off $\rho_{\infty}$ such that $e^{\rho_{\infty}}\propto \pi/4\epsilon$,  we find for $\beta=2\pi$ (computation details are given in the appendix)
\begin{equation}
\label{cftholo1}
\frac{\mbox{Vol}^{BTZ}(\Sigma)}{R^{2}}\simeq\frac{\pi V_1}{4\epsilon}+ V_1 t \qquad 2G^{BTZ}_{\lambda\lambda}\simeq \frac{\pi V_1}{4\epsilon}+V_1 t
\end{equation}
for the late time limit $(t\gg\beta~~\mbox{or}~~\kappa_{\ast}\rightarrow \pi/4)$  
as well as 
\begin{equation}
\label{cftholo2}
\frac{\mbox{Vol}^{BTZ}(\Sigma)}{R^{2}}\simeq\frac{\pi V_1}{4\epsilon}+ \frac{2}{\pi} V_1 t^2 \qquad 2G^{BTZ}_{\lambda\lambda}\simeq \frac{\pi V_1}{4\epsilon}+\frac{2}{3} V_1 t^2
\end{equation}
for the early time limit $(t\rightarrow 0~~\mbox{or}~~\kappa_{\ast}\rightarrow 0)$, recovering the dual results for the BTZ black hole and (in both cases) 
 the holography relation 
 \begin{equation}
\label{holgraphic1}
2G^{BTZ}_{\lambda\lambda}\simeq   n^{ \textsc{BTZ}}_1 \frac{\mbox{Vol}^{BTZ}(\Sigma)}{R^2}
\end{equation}
claimed in \cite{gravitydual}.   
The above results have been obtained through computations expressed in details in the appendices (\ref{appendix1}, \ref{appendix2}).

For the geon space the computation is similar, except that the symmetry of figure \ref{fig:M1} is
absent in figure \ref{fig:M2}
and so
the volume $\mbox{Vol}(\Sigma)$ is reduced to  half the  value for the BTZ case $(\mbox{Vol}^{ \textsc{geon}}(\Sigma) =
\mbox{Vol}^{ \textsc{BTZ}}(\Sigma)/2)$ 
and the quantum information metric \eqref{Geon-infomet} is twice the BTZ value.  We therefore obtain
\begin{equation}
\label{geonholo}
2G^{ \textsc{geon}}_{\lambda\lambda}   = n^{ \textsc{geon}}_1 \frac{\mbox{Vol}^{ \textsc{geon}}(\Sigma)}{R^2}
= 4   n^{ \textsc{BTZ}}_1 \frac{\mbox{Vol}^{ \textsc{geon}}(\Sigma)}{R^2} 
\end{equation}
which is consistent with the holographic relation  (\ref {holgraphic1}) but with a different factor.   This suggests that 
the coefficient $n_d$ is sensitive to the topology of spacetime. 

\section{Information Metric for Planar Black Holes and their Geon Counterparts}

We next consider  $d$-dimensional AdS Schwarzchild planar black holes,  with metric of
the general form
\begin{eqnarray}
\label{adsd+1}
ds^2&=&-f(r)dt^2+dr^2/f(r)+r^2dx^2_d\nonumber\\
f(r)&=&-M/r^{d-1}+r^2/l^2
\end{eqnarray}
which in turn becomes
\begin{eqnarray}
\label{planarbranes}
ds^2&=&\frac{1}{z^2}\big[-h(z)dt^2+\frac{dz^2}{h(z)}+dx^2_d\big]\nonumber\\
h(z)&=&1-z^{d+1} 
\end{eqnarray}
via the  change of variable $r=l/z$ and by setting $l=M=1$.

The quantum information metric for $\mbox{CFT}_{d+1}$ is given by
\begin{eqnarray}
\label{infoctfd}
&G^{SAdS_d}_{\lambda\lambda}&
=\frac{1}{2}\int_{\frac{\beta}{4}+\tau+\epsilon}^{\frac{3\beta}{4}-\tau-\epsilon}d\tau_2\int_{-\frac{\beta}{4}-\tau+\epsilon}^{\frac{\beta}{4}+\tau-\epsilon}d\tau_1 \int d^dx_1 \int d^dx_2\nonumber\\
&&\langle{\cal{O}}(x_1,\tau_1){\cal{O}}(x_2,\tau_2)\rangle_{SAdS_d}
\end{eqnarray}
with 
\begin{equation}\label{2-ptASd}
\langle{\cal{O}}(x_1,\tau_1){\cal{O}}(x_2,\tau_2)\rangle_{SAdS_d}=\frac{C_{12}}{|(\tau_1-\tau_2)^2+\sum^{d}_{1}(x_1-x_2)^2|^{\Delta}}
\end{equation}
where the coordinates $x$ are unwrapped (i.e. $x\in\mathbb{R}$) and  $C_{12}$ is a constant 
whose value will be subsequently fix. Upon carrying out the integration over $x_2$ the information metric takes the form 
\begin{eqnarray}
&&G^{SAdS_d}_{\lambda\lambda} = \frac{1}{2}\int_{\frac{\beta}{4}+\tau+\epsilon}^{\frac{3\beta}{4}-\tau-\epsilon}d\tau_2\int_{-\frac{\beta}{4}-\tau+\epsilon}^{\frac{\beta}{4}+\tau-\epsilon}d\tau_1\int d^dx_1\int dx\nonumber\\
&&\times\frac{\Gamma(\Delta-\frac{d-1}{2})}{\Gamma(\Delta)}\frac{C_{12} \pi^\frac{d-1}{2} (\pi/\beta)^{2(\Delta-\frac{d-1}{2}))}}{[\sinh^2(\frac{\pi x}{\beta})+\sin^2(\frac{\pi (\tau_1-\tau_2)}{\beta})]^{\Delta-\frac{d-1}{2}}} 
\end{eqnarray}
Choosing $C_{12}=\frac{\Gamma(\Delta)}{\pi^\frac{d-1}{2}\Gamma(\Delta-\frac{d-1}{2})}$, for an exactly marginal deformation $\Delta=d+1$ \cite{gravitydual} the information metric can be put into the form 
\begin{eqnarray}
G^{SAdS_d}_{\lambda\lambda}&=&\frac{1}{2}\int_{\frac{\beta}{4}+\tau+\epsilon}^{\frac{3\beta}{4}-\tau-\epsilon}d\tau_2\int_{-\frac{\beta}{4}-\tau+\epsilon}^{\frac{\beta}{4}+\tau-\epsilon}d\tau_1\int d^dx_1\int dx\nonumber\\
&&\times \frac{(\frac{\pi}{\beta})^{d+3}}{[\sinh^2(\frac{\pi x}{\beta})+\sin^2(\frac{\pi (\tau_1-\tau_2)}{\beta})]^{(d+3)/2}}
\end{eqnarray} 
 which is quite similar to the $\mbox{CFT}_2$ case, the only difference is that the integral over $x_1$ is taken over a $d$-dimensional spacetime. Using the expression
\begin{eqnarray}
&&\int^\infty_{-\infty}\frac{dx}{[\sinh^2 x+\sin^2 t]^n}\nonumber\\
&=& 2^{2n-1}\sin^n2t \beta (n,n)F_1(n,n,n, 2n, 1+1/a, 1+1/b)\nonumber
\end{eqnarray}
with ~ $a=\cot 2t + i$~ and~ $b=\cot 2t - i$ , we are led finally for the late time limit  ($\tau >>\beta$~ and~ $d$ odd) to
\begin{equation}
G^{SAdS_d}_{\lambda\lambda}\simeq \frac{V_d}{\epsilon^d}+\bigg(\frac{2\pi}{\beta}\bigg)^{d+1} V_d\tau   
\end{equation}
and the early time limit ($\tau\rightarrow 0$~ and~ $d$ odd) to
\begin{equation}
G^{SAdS_d}_{\lambda\lambda}\simeq \frac{V_d}{\epsilon^d}+\bigg(\frac{2\pi}{\beta}\bigg)^{d+2} V_d\tau^2   
\end{equation} 
Here $V_d$ is an infinite $d$-dimensional volume. 

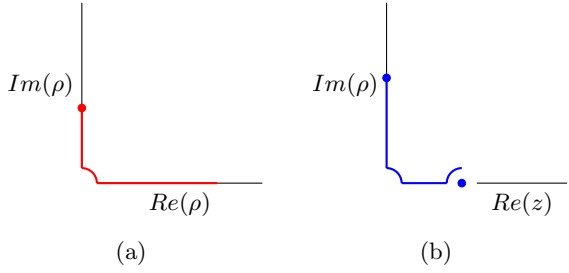
\begin{figure}
\begin{subfigure}[b]{0.22\textwidth}
\centering
\begin{tikzpicture}[scale=1]
\draw[-](0.2,0)--(2.4,0) node[midway,below] {$Re(\rho)$};
\draw[-](0,0.2)--(0,2.4)node[midway,left] {$Im(\rho)$};
\draw[red,fill=red] (0,1) circle (.05);
\draw[red, thick](0.2,0)--(1.8,0);
\draw[red, thick](0,0.2)--(0,1);
\draw[red, thick](0,0)(0:0.2)arc(0:90:0.2);
\end{tikzpicture}
\caption{}\label{fig:M4}
\end{subfigure}
\begin{subfigure}[b]{0.22\textwidth}
\centering
\begin{tikzpicture}[scale=1]
\draw[-](1.2,0)--(2.4,0) node[midway,below] {$Re(z)$};
\draw[-](0,0.2)--(0,2.4) node[midway,left] {$Im(\rho)$};
\draw[blue,fill=blue] (0,1.4) circle (.05);
\draw[blue,fill=blue] (1,0) circle (.05);
\draw[blue, thick](0.2,0)--(0.8,0);
\draw[blue, thick](0,0.2)--(0,1.4);
\draw[blue, thick](0,0) (0:0.2)arc(0:90:0.2);
\draw[blue, thick](1,0) +(180:0.2)arc(180:90:0.2);
\end{tikzpicture}
\caption{}\label{fig:M5}
\end{subfigure}
\caption{(a) The contour chosen on the complex plane $\rho$  to compute $\mbox{Vol}(\Sigma_{max})$ for the BTZ black hole. $\kappa_{\ast}$ is the point on the imaginary axis. (b) Represents the contour chosen on the complex plane $z$ for the computation of $\mbox{Vol}(\Sigma_{max})$ for the AdS Schwarzschild planar black hole. $z_{\ast}$ on the imaginary axis is at $\sqrt{2}z_0$ and at $z_0$ on the real axis.}
\label{fig:M6}
\end{figure}

 Proceeding as before, we define a hypersurface  $z=z(t)$. Its volume is  
\begin{equation}
\frac{\mbox{Vol}^{SAdS_{d+2}}(\Sigma)}{R^{d+1}}=V_d\int\frac{dt}{z^{d+1}\sqrt{h(z)}}\sqrt{{\dot{z}}^2-{h(z)}^2}
\end{equation}
Following the same steps as in the previous section, we are led to a maximum volume (with $z_{\ast}$ the value of $z$ such that $\partial z/\partial t=0$)
\begin{eqnarray}
\frac{\mbox{Vol}^{SAdS_{d+2}}(\Sigma)}{R^{d+1}}&=&2V_d\int\frac{dz}{z^{d+1}\sqrt{h}\sqrt{1-(z/z_{\ast})^{2(d+1)}(h_{\ast}/h)}}\nonumber\\
t&=&\int\frac{dz}{h\sqrt{1-(z_{\ast}/z)^{2(d+1)}(h/h_{\ast})}}
\label{volbtzplan}
\end{eqnarray}
In the late time limit $(t\gg\beta~~\mbox{or}~~z_{\ast}\rightarrow 2^{\frac{1}{d+1}}z_0)$, we obtain
\begin{eqnarray}
\frac{\mbox{Vol}^{SAdS_{d+2}}(\Sigma)}{R^{d+1}}&=&2V_d\big[-i\int^{2^{\frac{1}{d+1}}z_0}_{0}\frac{dz}{z^{d+1}\big[1+\frac{1}{2}(\frac{z}{z_0})^{d+1}\big]}\nonumber\\
&+&\int^{z_0}_{0}\frac{dz}{z^{d+1}\big[1-\frac{1}{2}(\frac{z}{z_0})^{d+1}\big]}\big]\nonumber\\
t&=&\frac{1}{2z^{d+1}_{0}}\big[-i\int^{2^{\frac{1}{d+1}}z_0}_{0}\frac{z^{d+1}dz}{(1+\frac{z^{d+1}}{z^{d+1}_0})\big[1+\frac{1}{2}(\frac{z}{z_0})^{d+1}\big]}\nonumber\\
&+&\int^{z_0}_{0}\frac{z^{d+1}dz}{(1-\frac{z^{d+1}}{z^{d+1}_0})\big[1-\frac{1}{2}(\frac{z}{z_0})^{d+1}\big]}\big]
\end{eqnarray}
 with the integration contour given in \ref{fig:M5}.
Integrating out this expressions, we find the same behaviour as in the previous case  (with $z_0=\beta/{2\pi}$)
 \begin{eqnarray}
&&\frac{\mbox{Vol}^{SAdS_{d+2}}(\Sigma)}{R^{d+1}}\simeq \frac{ V_d}{d \epsilon^d}+\bigg(\frac{2\pi}{\beta}\bigg)^{d+1}V_d t\nonumber\\ &&2G^{SAdS_{d+2}}_{\lambda\lambda}\simeq \frac{V_d}{\epsilon^d}+\bigg(\frac{2\pi}{\beta}\bigg)^{d+1}V_d t
\end{eqnarray}
When looking at the early time limit $(t\rightarrow 0~~\mbox{or}~~z_{\ast}\rightarrow z_0)$, we have 
\begin{eqnarray}
&&\frac{\mbox{Vol}^{SAdS_{d+2}}(\Sigma)}{R^{d+1}}=2V_d\big[\int^{z_0}_{0} \frac{1}{z^{d+1}\sqrt{h}}dz+\frac{1}{2}\frac{h_{\ast}}{z^{2(d+1)}_{\ast}}\int^{z_0}_{0}\frac{z^{d+1}}{\sqrt{h^3}}dz\big]\nonumber\\
&&t=-i\frac{\sqrt{h_{\ast}}}{z^{d+1}_{\ast}}\big[\int^{z_0}_{0}\frac{z^{d+1}}{\sqrt{h^3}}dz+\frac{1}{2}\frac{h_{\ast}}{z^{2(d+1)}_{\ast}}\int^{z_0}_{0}\frac{z^{3(d+1)}}{\sqrt{h^5}}dz\big]\nonumber\\
&&
\end{eqnarray}
after integrating 
\begin{eqnarray}
&&\frac{\mbox{Vol}^{SAdS_{d+2}}(\Sigma)}{R^{d+1}} \simeq \frac{V_d}{d\epsilon^d}+\frac{(d+1)^2}{2\beta(\frac{1}{2},\frac{1}{d+1})}\bigg(\frac{2\pi}{\beta}\bigg)^{d+2}V_d t^2\nonumber\\ &&2G^{SAdS_{d+2}}_{\lambda\lambda}\simeq \frac{V_d}{\epsilon^d}+ \bigg(\frac{2\pi}{\beta}\bigg)^{d+2}V_d t^2
\end{eqnarray}
which is similar to the previous case with ($d=1$). Hence we obtain
\begin{equation}
 2 G^{SAdS_d}_{\lambda\lambda}= n^{SAdS}_d\frac{\mbox{Vol}^{SAdS}(\Sigma_{max})}{R^{d+1}}
\end{equation}
which is an extension of  \eqref{holgraphic1} to $d$-dimensions.

Turning now to the geon, the identification \eqref{geontransf}   acts on the same pair of coordinates with the other coordinates remaining invariant. For the holographic computations in $(d+2)$-dim the volume $V_1$ in $(1+2)$-dim is replaced by the (infinite) volume $V_d$ associated with \eqref{planarbranes}.  Likewise, 
we find that the generalization of the two-point function \eqref{2-ptASd} for the geon consists of two contributions of equal value,  and so we obtain 
\begin{equation}
G^{SAdSgeon_d}_{\lambda\lambda}=2   G^{SAdS_d}_{\lambda\lambda}
\end{equation} 
Putting all these aforementioned results together, we find  
\begin{eqnarray}
\label{intro3}
2 G^{SAdSgeon_d}_{\lambda\lambda}&=&n^{SAdSgeon}_d\frac{\mbox{Vol}^{geon}(\Sigma_{max})}{R^{d+1}}\nonumber\\
&=&4n^{SAdS }_d\frac{\mbox{Vol}^{geon}(\Sigma_{max})}{R^{d+1}}
\end{eqnarray} 

for the planar Schwarzschild AdS black hole.

\section{Conclusion}
 
By investigating both bulk and boundary contributions for the BTZ black hole and the $d$-dimensional planar Schwarzschild AdS black hole, we have found that the relation \eqref{intro1} holds for their geon
counterparts apart from a factor of 4. For each case, compared to its black hole counterpart 
 the information metric of the corresponding geon is twice as large and on the bulk side the maximum volume of a time slice for the geon is half as large.  We conclude that the relation \eqref{intro1}  (and thus the coefficient $n_d$)
 is sensitive to the topological structure of the spacetime, with $ n^{geon}_d = 4 n_d$
for the cases we have considered.   In this sense the information metric in the
 CFT is a `probe' of spacetime topology.
 
It would be interesting to explore this relationship further, extending the proposed relation
\eqref{intro1} to spacetimes of more interesting topology, including rotation, solitons, and more generalized geometries. 

\section*{Acknowledgements}
This work was supported by the Natural Sciences and Engineering Research Council of Canada.
 
\section*{APPENDIX}

Here we provide some detail on the computations computed in the bulk.
 
 Let us start with the BTZ black hole, in this case we obtain from \eqref{volbtz} for the late time limit
\begin{eqnarray}
\frac{\mbox{Vol}^{BTZ}(\Sigma)}{R^2V_1}&=&-\int^{\pi/4}_{0}\frac{\cos 2\kappa}{\sin\kappa}d\kappa +\int^{\rho_\infty}_{0}\frac{\cosh 2\rho}{\sinh\rho}d\rho + t\nonumber\\
t&=&\int^{\pi/4}_{0}\frac{d\kappa}{\sin\kappa\cos 2\kappa}-\int^{\rho_\infty}_{0}\frac{d\rho}{\sinh\rho\cosh 2\rho}\nonumber\\
&&
\end{eqnarray}
or, more explicitly
\begin{eqnarray}
\frac{\mbox{Vol}^{BTZ}(\Sigma)}{R^2V_1}&=&2\cosh\rho_\infty -\sqrt{2}-\log(1-\sqrt{2})+t\nonumber\\
t&=&\int^{\pi/4}_{0}\frac{d\kappa}{\sin\kappa\cos 2\kappa}-\int^{\rho_\infty}_{0}\frac{d\rho}{\sinh\rho\cosh 2\rho}\nonumber
\end{eqnarray}
Setting $e^{\rho_\infty}\simeq \pi/4\epsilon$, this can be approximated  to the asymptotic form 
\begin{equation}
\label{appendix1}
\frac{\mbox{Vol}^{BTZ}(\Sigma)}{R^{2}}\simeq \frac{\pi V_1}{4\epsilon}+V_1t + \cdots
\end{equation}
with $t\sim -\log\epsilon$, 
which is congruent with the results we got from the CFT computations. We can also see that the closer the parameter $\epsilon$ get to $0$ the greater is the time $t$. 
\vskip 5pt When looking at the early time limit, we find from \eqref{volbtz} that
\begin{eqnarray}
\frac{\mbox{Vol}^{BTZ}(\Sigma)}{R^2V_1}&=&2\int^{\rho_\infty}_{0}\cosh\rho d\rho+ \sin ^2 2\kappa_{\ast}\int^{\rho_\infty}_{0}\frac{d\rho}{\sinh\rho\sinh 2\rho}\nonumber\\
t&=&-\sin 2\kappa_{\ast}\int^{\rho_\infty}_{0}\frac{d\rho}{\sinh\rho\sinh 2\rho}
\end{eqnarray}
which can also take the form 
\begin{equation}
\label{appendix2}
 \frac{\mbox{Vol}^{BTZ}(\Sigma)}{R^{2}}\simeq \frac{\pi V_1}{4\epsilon}+\frac{2}{\pi}t^2 
\end{equation}
where now $t\simeq -(\frac{1}{\epsilon}-\frac{\pi}{2})\sin(2\kappa_{\ast})$.
Here we notice that the term in $1/\epsilon$ can be eliminated by means of renormalization of the time $t$ in \eqref{appendix2}, rendering the latter finite and converging to $0$ when $\kappa_{\ast}$ approaches $0$. 

 Concerning  the AdS Schwarzschild planar black hole, we obtain  in the late time limit 
\begin{eqnarray}
\frac{\mbox{Vol}^{SAdS_{d+2}}(\Sigma)}{R^{d+1}}&=&2\frac{V_d}{z_0^d}\big[-\frac{i}{2^\frac{d}{d+1}}\int^{1}_{\epsilon}\frac{dx}{x^{d+1}\big[1+x^{d+1}\big]}\nonumber\\
&+&\int^{1-\epsilon}_{\epsilon}\frac{dx}{x^{d+1}\big[1-\frac{1}{2}x^{d+1}\big]}\big]\nonumber\\
\end{eqnarray}
from \eqref{volbtzplan}, 
or explicitly
\begin{eqnarray}
\frac{\mbox{Vol}^{SAdS_{d+2}}(\Sigma)}{R^{d+1}}&=&2\frac{V_d}{z_0^d}\big[-2^\frac{d}{d+1}i \int^{1}_{\epsilon}\frac{x^{d+1}dx}{(1+x^{d+1})\big[1+2x^{d+1}\big]}\nonumber\\
&-&i\int^{1}_{\epsilon}\frac{\beta_1 dx}{1+x^{d+1}} \nonumber\\
&+&\int^{1-\epsilon}_{\epsilon}\frac{x^{d+1}dx}{(1-x^{d+1})\big[1-\frac{1}{2}x^{d+1}\big]}\big]\nonumber\\
&+&\int^{1-\epsilon}_{\epsilon}\frac{\beta_2 dx}{1-\frac{1}{2}x^{d+1}}\nonumber\\
t&=&z_0\big[- 2^\frac{d}{d+1}i\int^{1}_{\epsilon}\frac{x^{d+1}dx}{(1+x^{d+1})\big[1+2x^{d+1}\big]}\nonumber\\
&+&\int^{1-\epsilon}_{\epsilon}\frac{x^{d+1}dx}{(1-x^{d+1})\big[1-\frac{1}{2}x^{d+1}\big]}\big]
\end{eqnarray}
with 
\begin{eqnarray}
\beta_1&=&\frac{2^\frac{1}{d+1}(1+2x^{d+1})-2^\frac{d}{d+1}x^{2d+2}}{x^{d+1}(1+2x^{d+1})}\nonumber\\
\beta_2&=&\frac{2-2x^{d+1}-x^{2d+2}}{x^{d+1}(1-x^{d+1})}\nonumber
\end{eqnarray}
This can be finally put into the asymptotic form 
\begin{equation}
\label{appendix3}
\frac{\mbox{Vol}^{SAdS_{d+2}}(\Sigma)}{R^{d+1}} \simeq  \frac{V_d}{d\epsilon^d}+\frac{V_d t}{z_0^{d+1}} 
\end{equation}
with $t \sim\epsilon^\frac{1}{d+1}{_2}F_1\big[\frac{-1}{d+1}, \frac{-1}{d+1}, \frac{d}{d+1}, \frac{1}{(d+1)\epsilon}\big]$.
The early time limit reads
\begin{eqnarray}
\frac{\mbox{Vol}^{SAdS_{d+2}}(\Sigma)}{R^{d+1}}&=&2\frac{V_d}{z_0^d}\big[\int^{1-\epsilon}_{\epsilon}\frac{dx}{x^{d+1}\sqrt{1-x^{d+1}}}\nonumber\\
&+&\frac{1}{2}h_{\ast}\int^{1-\epsilon}_{\epsilon}\frac{x^{d+1}dx}{\sqrt{(1-x^{d+1})^3}}\big]\nonumber\\
t&=&-iz_0\sqrt{h_{\ast}}\int^{1-\epsilon}_{\epsilon}\frac{x^{d+1}dx}{\sqrt{(1-x^{d+1})^3}}
\end{eqnarray}
Hence 
\begin{equation}
\label{appendix4}
\frac{\mbox{Vol}^{SAdS_{d+2}}(\Sigma)}{R^{d+1}}\simeq 2V_d\big[\frac{1}{d\epsilon^dz_0^d}+\frac{(d+1)^2}{4\beta (\frac{1}{2},\frac{1}{d+1})} \frac{t^2}{z_0^{d+2}}\big]
\end{equation}
where now $t \simeq -iz_0\sqrt{h_{\ast}}\big[\frac{2}{\sqrt{(d+1)^3\epsilon}}-\frac{2}{(d+1)^2}\beta (\frac{1}{2}, \frac{1}{d+1})\big]$.
The term in $1/\sqrt{\epsilon}$ can be removed by renormalizing this expression so that the time $t$ goes to $0$ when $z_{\ast}=z_0$.

\end{document}